\documentclass[preprint,12pt]{elsarticle}
\usepackage{amsmath,amsfonts,amssymb}
\usepackage{subfigure}
\usepackage{pdfsync}
\usepackage[para]{threeparttable}
\usepackage{hyperref}
\usepackage[version=3]{mhchem}
\biboptions{sort&compress}

\newcommand{\f}{\frac}

\newcommand{\suml}{\sum\limits}

\newcommand{\intl}{\int\limits}

\graphicspath{{D:/localtexmf/Mytex/LGH/JMolLiq/SurfTens/figures/}}
 \newcommand*{\OrigAA}{}
\let\OrigAA\AA

\renewcommand*{\AA}{%
  {\fontfamily{ptm}%
  \selectfont%
  \OrigAA%
  \selectfont}%
}
\journal{Journal of Molecular Liquids}
\begin{document}
\begin{frontmatter}
\title{Surface tension of molecular liquids: Lattice gas approach}
\author[am]{A.~Maslechko}
\ead{anastasiia.maslechko@ntnu.no}%
\author[kg]{K.~Glavatskiy}
\ead{}%
\author[vlk]{V.L. Kulinskii\corref{cor1}}
\ead{kulinskij@onu.edu.ua}%
\cortext[cor1]{Corresponding author}
%
\address[am]{Department of Chemistry, Norwegian University of Science and Technology, Hogskoleringen 5, N-7491 Trondheim, Norway}
\address[kg]{School of Chemical Engineering, The University of Queensland, Australia}
\address[vlk]{Department of Theoretical
Physics, Odessa National University, Dvoryanskaya 2, 65082 Odessa, Ukraine}

\begin{abstract}
The approach of global isomorphism between the fluid and the Ising model is applied to obtain an expression for the surface tension of the Lennard-Jones fluid on the basis of the information about the Ising model. This is done in a broad interval of temperatures along the phase coexistence, and is valid both in 2D and 3D. The relation between the critical amplitudes of the surface tension of the fluid and the Ising model is derived in the vicinity of the critical point. The obtained theoretical estimates agree well with the literature results for the surface tension. The methodology is demonstrated for the 2D LJ fluid on the basis of the exact solution of the 2D Ising model and is tested for the 3D LJ fluid. As a result, an expression for the surface tension without any fitting parameter is derived.
\end{abstract}
\end{frontmatter}

\section{Introduction}
The surface tension is one of the most vivid properties of a fluid. Theoretical descriptions, which employ methods of homogeneous statistical mechanics, are challenging to use due to lack of symmetry at the phase boundary, where the system is spatially inhomogeneous. The seminal square gradient approximation of van der Waals \cite{eos_vdwsqgrad_zpch1894} (see also \cite{crit_vdwscaling_physica1974} for historical survey) and more sophisticated approaches \cite{book_widomrowlinson,liq_ljdensprofile_jcp1982,crit_densityprof_molphys1985}, along with the Principle of Corresponding State (PCS) are the basis for understanding of the universal regularities at the phase coexistence, which are valid for a very broad class of fluids and for a broad region of fluid states, including the vicinity of the critical point \cite{book_ll5_en}.

Two remarkable empirical relations for fluids are known for more than a century.
One of them is the law of rectilinear density diameter (LRD) \cite{crit_diam1}, which states that if $n_{l},n_{g}$ are the densities of corresponding phases along the binodal, and $T$ is the temperature, then:
\begin{equation}
  n_{d}/n_{c} = \f{n_{l}+n_{g}}{2\,n_c} =  1+A\,
\left(\, 1-T/T_c \,\right)\,,
\label{rdl}
\end{equation}
where $A$ is a constant. The other linear relation is the Zeno line (ZL) linearity \cite{eos_zenobenamotz_isrchemphysj1990}, which states that the unit compressibility curve $Z \equiv P/(n\,T) = 1$ is described by a linear equation:
\begin{equation}\label{z1}
\f{T}{T_{Z}}+\f{n}{n_Z} = 1\,,
\end{equation}
where the parameters $T_Z$ and $n_Z$ are determined from the expressions for the virial coefficients~\cite{eos_zenobenamotz_isrchemphysj1990}:
\begin{equation}\label{tbnb}
  B_2(T_{Z}) = 0\,,\quad n_Z= \f{ T_Z }{B_3\left(\,T_Z\,\right)}\,\left. \f{dB_2}{dT}\right|_{T= T_Z}\,.
\end{equation}
For the van der Waals equation of state ZL is fulfilled trivially and is known as the Batschinski law \cite{eos_zenobatschinski_annphys1906}.

Recently, the classical PCS has gained a new development by analyzing universality in the behavior of the density binodal rectilinear diameter and the Zeno-line linearity \cite{eos_zenoholleran_jpc1969,eos_zenobenamotz_isrchemphysj1990}, which are observed in quite versed fluid systems, which span standard molecular liquids, various model fluids with the pair interaction potentials of Mie-class, and liquid metals \cite{eos_zenoapfelbaum_jpchemb2008,eos_zenoapfelbaum1_jpcb2009,eos_zenoline_potentials_jcp2009}.
These findings were casted in a simple geometrical picture of the global isomorphism (GI) between the fluid and the Ising model, where the asymmetric fluid binodal is symmetrized to resemble the symmetric binodal of the Ising model \cite{eos_zenome_jphyschemb2010,crit_globalisome_jcp2010}. Subsequently, the concept of the global isomorphism between the states of the Lennard-Jones (LJ) fluid and the Ising model was elaborated \cite{eos_zenogenpcs_jcp2010}. The developed global isomorphism incorporates PCS in a natural way. Indeed, molecular liquids, which show linearity of the rectilinear diameter as well as Zeno-line linearity, belong to a single class, which has the property that the fluid part of the phase diagram is topologically isomorphic to the phase diagram of the Ising model \cite{eos_zenozcassocme_jcp2014}.

The aim of this paper is to add to the results of recent work \cite{eos_zenosurftensus_jphchemc2016} by considering the consequences of the global isomorphism for the critical amplitudes of the surface tension and the correlation length of the LJ fluid as well as the interfacial width. These relations can be verified in computer simulations. We use available data to verify some of them.

The paper is organized as follows. In Section~\ref{sec_zenoiso} we outline the main ideas of the global isomorphism between the molecular fluid with the LJ type of interaction and the lattice gas. In Section~\ref{sec_st} the critical amplitude of the surface tension for the LJ fluid is derived and the compared with the results from literature. In Section~\ref{sec_width} we discuss the effective interfacial width and its relation to the correlation length. Concluding remarks and discussion of the future ideas are presented in the Conclusion Section.

\section{Global isomorphism and the Zeno-line}\label{sec_zenoiso}
Consider the lattice gas model, which is described by the Hamiltonian  $H_{latt}$ (see e.g. \cite{book_baxterexact}):
\begin{equation}\label{hamlg}
  H_{latt} = -\varepsilon\,\suml_{\langle i,j \rangle}\,n_{i}\,n_{j} - \mu\,\suml_{i}\,n_{i}
\end{equation}
where $\varepsilon$ is the energy of the nearest cite-cite attraction, $\mu$ is the chemical potential, and the repulsive part is modeled by the restriction for the occupation number of a cite $n_{i} = 0,1$. The lattice gas model is isomorphic to the Ising model \cite{book_baxterexact} with the interaction constant $J$ related to the energy of the nearest cite-cite attraction as \cite{book_huang_statmech} $\varepsilon = 4\,J$. We will use the  names "`Ising model"' and "`lattice gas model"' interchangeable, emphasizing the difference, where necessary.

Let $x= \langle n_i\rangle$ be the density of the lattice gas and $\tilde{t} = t/t_c$ be the temperature of the lattice gas normalized by its critical temperature value $t_c$.
Then the global isomorphism between the Ising model (or, equivalently, the lattice gas model) and the LJ fluid with the density $n$ and the temperature $T$ is represented by the following projective transformation \cite{eos_zenome_jphyschemb2010,eos_zenomeglobal_jcp2010}:
\begin{equation}\label{projtransfr_my}
  n(x,\tilde{t}) =\, n_{*}\,\f{x}{1+z\,\tilde{t}}\,,\quad
  T(\tilde{t}) =\, T_{*}\,\f{z\, \tilde{t}}{1+z\,\tilde{t}}
\end{equation}
and the corresponding inverse transformation
\begin{equation}\label{projtransfrinv_my}
 x =  \frac{n}{n_{*}}\frac{T_*}{T_*-T}\,,\quad
  \tilde{t} =\frac{1}{z}\,\frac{T}{T_*-T}
\end{equation}
where the parameter $z$ is given by
\begin{equation}\label{zttc_my}
z = \f{T_c}{T_{*}-T_c}
\end{equation}
and the parameters $n_{*},T_{*}$ are the ones, which determine the linear zeno-element
\begin{equation}\label{eq_zenomy}
  \f{n}{n_{*}}+\f{T}{T_{*}} = 1
\end{equation}
The temperature $T_{*}$ is related to the Boyle point in van der Waals (vdW) approximation:
\begin{equation}\label{tbvdwmy}
  B^{vdW}_2(T_{*}) = 0\,,\quad  T_{*} =  T^{(vdW)}_Z  = \f{a}{b}
\end{equation}
where $a = -2\pi\,\intl_{r_*}^{+\infty}\Phi_{attr}(r)\,r^2\,dr$ and $b = \f{2\pi}{3}\,d^{3}$, where $d$ is the particle diameter. Furthermore, the density $n_{*}$ represents the high density state \cite{eos_zenobenamotz_isrchemphysj1990,eos_zeno_jphyschem1992}:
\begin{equation}\label{nbvdwmy}
n_*= \f{ T_* }{B_3\left(\,T_*\,\right)}\,\left. \f{dB_2}{dT}\right|_{T= T_*}\,, \quad n_{*}\approx 1/b
\end{equation}
For the LJ fluid we use conventional dimensionless units for the temperature $T\to T/\Phi_0$, density $n\to n d^3$.

Note, that for the Ising model the law of rectilinear diameter \eqref{rdl} is fulfilled trivially because of the spin-flop symmetry. In that case the Zeno-line is the line $x=1$ where $x=N/\mathcal{N}$ is the molar fraction of $N$ occupied sites on a lattice with total $\mathcal{N}$ sites.

The interaction potential of the LJ fluid is
\begin{equation}\label{lj}
  \Phi_{LJ}(r) = 4\,\Phi_0\,
\left(\, \left(\, \f{d}{r}  \,\right)^{12} - \left(\, \f{d}{r}  \,\right)^{6}\,\right)
\end{equation}
Since its attractive part has asymptotic behavior $\phantom{1}\sim - r^{-6}$, it is possible to sow that
\begin{equation}\label{zd}
  z = \frac{D}{6}
\end{equation}
where $D$ is the spatial dimension and 6 is the power the attractive part of the potential \cite{eos_zenomeglobal_jcp2010}.

The relation between the parameter $\Phi_0$ and the lattice gas interaction parameter $\varepsilon$ in Eq.~\eqref{hamlg} is independent of the spatial dimension \cite{book_huang_statmech, book_baxterexact}: $\Phi_0 = \varepsilon$. Furthermore, $T_{*} = 2$ in two dimensions (2D) and $T_{*} = 4$ in three dimensions (3D) \cite{eos_zenomeglobal_jcp2010}. A simple estimate based on Eq.~\eqref{projtransfr_my} leads to the following values of the critical temperatures of the LJ fluid:
\begin{equation}\label{tcljd2d3}
T^{{(2D)}}_{c} =1/2\,,\quad T^{{(3D)}}_{c} =4/3
\end{equation}
These values are in good agreement with the results of computer simulations \cite{crit_lj2dim_jcp1991,crit_simulreviewpanag_molphys1992}.

It is important to emphasize that linear behavior of the binodal diameter and of the Zeno-line have much broader validity than the standard formulation of the PCS based on a simple scaling law of the interaction potentials \cite{eos_zenoherschbachktrans_jpcb2013}. In particular, the behavior of the surface tension also reveals universal features \cite{crit_surftenspcs_jcp2012} and can be understood in the context of the global isomorphism formulation of the PCS. Thus, the Zeno-line regularity has been used in the studying the surface tension of molecular liquids \cite{eos_zenosurftensus_jphchemc2016}. This approach is based on the relation between thermodynamic potentials of the lattice gas and the LJ fluid stated in \cite{eos_zenomeunified_jphyschemb2011}. In particular, Eq.~\eqref{projtransfr_my} follows from the equality between the grand thermodynamic potential of the LJ fluid, $\Psi \equiv P(\mu(h),T(t))\,V - \gamma_{LJ}\,\mathcal{A}$, and the one of the isomorphic lattice model, $G(h,t) \equiv \mathcal{N}\,g(h,t)-\gamma_{latt}\,\Sigma$:
\begin{equation}\label{potentials_my}
\Psi = G
\end{equation}
where $P$ is the pressure and $\mu$ is the chemical potential of the LJ fluid, $V$ is the fluid volume, while $h$ is the field variable conjugated to $x$, $\mathcal{N}$ is the number of sites in the lattice of the lattice gas.

The strength of the global isomorphism approach can also be demonstrated by considering the liquid-vapor equilibrium for the 2D LJ fluid \cite{crit_globalisome_jcp2010} on the basis of the Onsager's exact solution. Before doing that, we should note that computer simulations do not always agree well with the existing theoretical approaches \cite{eos_2dljabraham_physrep1981, crit_lj2dim_jcp1991, crit_2dsurftens_jcp1996, liq_2dsurftens_jcp2009}.The use of projective transformation \eqref{projtransfr_my} to map the binodal of the Ising model to the liquid-vapor coexistence curve gives more adequate estimate for the critical point and correctly reproduces the behavior of the binodal near the critical point \cite{crit_globalisome_jcp2010}. This is a consequence of a small value of the order parameter critical index ($\beta=1/8$) and the global character of its power law behavior. Also Eq.~\eqref{potentials_my} along with Eq.~\eqref{projtransfr_my} leads to a simple relation between the  critical compressibility factors, $Z_c = P_c/(n_c\,T_c)$ of the lattice gas (Ising model) and the LJ fluid \cite{eos_zenozcme_jcp2013}:
\begin{equation}\label{zc_my}
  Z_{c,\,LJ} =\f{(1+z)^2}{z}\,\f{t_c}{T_*}\, Z_{c,\,latt}
\end{equation}
where $t_c$ and $T_c$ are the critical temperatures of the lattice gas and the LJ fluid respectively. In the 3D case $z=1/2$, $Z_{c,\,latt} = 0.221$ for cubic lattice \cite{crit_dombsykes_prc1956} and Eq.~\eqref{zc_my} leads to:
\begin{align}\label{zc_my_lj3d}
  Z_{c,\,LJ}\approx 1.27\,Z_{c,\,latt} = 0.281
\end{align}
This value agree very well with the known values of $Z_c$ for the noble fluids \ce{Ar}, \ce{Kr}, \ce{Xe}, or the fluids like \ce{CO2}, which are usually considered as the canonical examples of the LJ fluid.

These results encourage us to apply the global isomorphism relations to the surface tension. Earlier \cite{eos_zenosurftensus_jphchemc2016} it was shown how to use the results for the surface tension of the Ising model in order to get the surface tension of the molecular fluids in a wide temperature region.

\section{Relation between critical amplitudes for surface tension of the lattice gas and the LJ liquid}\label{sec_st}
Starting from the relation \eqref{potentials_my} between thermodynamic potentials of the LJ fluid and the lattice gas, and taking into account that the linearities \eqref{rdl},\eqref{z1} in the bulk phase and in the surface decouple, according to \eqref{potentials_my}, we establish the correspondence between the surface tensions of the LJ fluid and the lattice model:
\begin{equation}\label{st_iso}
\gamma_{LJ}(T)= \gamma_{latt}(t(T))
\end{equation}
provided that geometrical sizes of the corresponding systems are the same.

In the following analysis we will use the dimensionless units for the surface tension: $\sigma_{latt}=\gamma_{latt} l^{D-1}/J$, where $J$ is the interaction constant of the Ising model (and not the lattice gas model) and $l$ is the lattice spacing. The surface tension of the LJ fluid is measured in corresponding units: $\sigma_{LJ}=\gamma_{LJ} d^{D-1}/\Phi_0$, where $\Phi_0$ and $d$ are the parameters of the LJ potential ~\eqref{lj}.

In the context of the global isomorphism, Eq.~\eqref{st_iso} allows one to extract the information about the surface tension of the LJ fluid on the basis of the information for the corresponding lattice model. In particular, for the surface tension of the 2D LJ fluid on the basis of the Onsager's solution \cite{crit_onsager_pr1944} we have:
\begin{equation}\label{st_2disingfit}
\sigma^{(2D)}_{LJ}(T) = \frac{1}{4}\,\left(2 + t(T)\,\ln
\left(\,\tanh\,\frac{1}{t(T)} \,\right)\right)\,,
\end{equation}
with
\[t(T)/t_{c} = 3\,\frac{T}{2-T}\]

Another immediate consequence of \eqref{st_iso} concerns the critical amplitude $s^{(0)}$ of the surface tension:
\begin{equation}\label{critamplit_st}
  \sigma = s^{(0)}\,|\tau|^{(D-1)\,\nu}+\ldots\,,\quad \tau = 1-T/T_c
\end{equation}
The relation between the critical amplitudes for the surface tension of the corresponding systems is as follows:
\begin{equation}\label{critamplit_st_relat}
s^{(0)}_{LJ} = \frac{1}{4}\,s^{(0)}_{Is}\,(1+z)^{(D-1)\,\nu}
\end{equation}
The factor $1/4$ in Eqs.~\eqref{st_2disingfit} and \eqref{critamplit_st_relat} appears due to the definition of the interaction constant of the Ising model.

\begin{figure}
  \centering
  \includegraphics[scale=0.43]{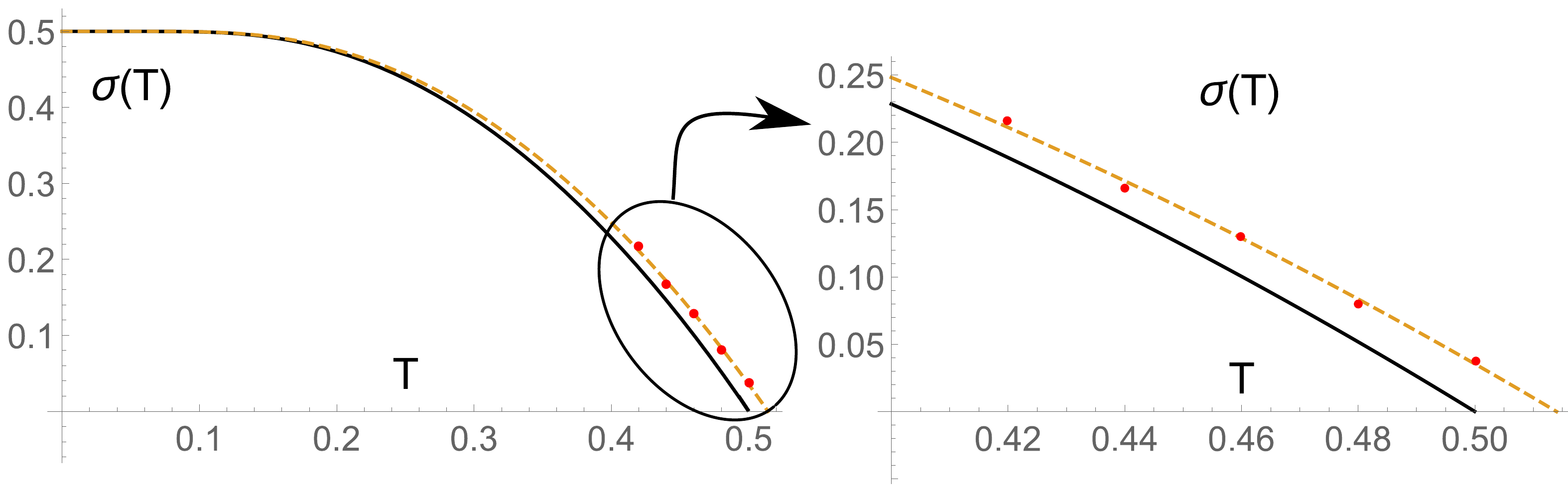}
  \caption{Dashed curve: the result of fitting Eq.~\eqref{st_2disingfit} to the data of \cite{liq_2dsurftens_jcp2009}(points) with $T_{*}= 2.08$, $1/z =  3.053$, $T_{c} = 0.514$. Solid curve: Eq.~\eqref{st_2disingfit} with the parameter values predicted by global isomorphism approach $z=1/3,\, T_{*} = 2,\, T_{c}=0.5$.}\label{fig_sigma2dfit}
\end{figure}

We check the validity of our approach by applying Eq.~\eqref{st_2disingfit} to the results of molecular simulations \cite{liq_2dsurftens_jcp2009} using $T_{*}$ and $z$ as the fitting parameters. The result of our fitting is shown in Fig.~\ref{fig_sigma2dfit}. We see, that the simulation data agree quite well with the predictions of the global isomorphism approach. In addition, the value of the critical amplitude in 2D case, $s^{(0)}_{LJ} \approx 1.318$ which follows from the simulations \cite{liq_2dsurftens_jcp2009}, is also in good agreement with the theoretical value, $s^{(0)}_{LJ} = 4/3$. We can conclude, that application of the global isomorphism to the surface tension of the 2D LJ fluid gives much better results than the other theoretical approaches used to analyze the results of the simulations \cite{liq_2dsurftens_jcp2009}, since they fail to reproduce both the binodal and the surface tension data.

The relation \eqref{critamplit_st_relat} is the direct consequence of the global isomorphism approach and can be useful in comparison between the results for the lattice and the fluid systems. The results for the critical amplitude of the surface tension are summarized in Table~\ref{tab_stamplit}. The typical values of the amplitude obtained from the literature are presented for the comparison. Note, that these values are scattered in a broad interval $2.1\le s^{(0)}_{LJ}\le 2.94$. This uncertainty is due to inaccuracy in fixing the position of the critical point, which is caused by truncation of the LJ potential. This is especially crucial for the surface properties of a spatially inhomogeneous fluid rather than for bulk thermodynamic properties of a homogeneous state \cite{liq_ljtailcorr_molphys1995, eos_ljsurfacetension_jcp1999, liq_nucleat2d_jcp2008}. At the same time, it seems that regardless of the truncation radius, the linearity of the rectilinear diameter and of the Zeno-line are observed in computer simulations for a broad set of potentials, including Mie-potentials \cite{eos_zenoline_potentials_jcp2009}.
\begin{table}[hbt!]
  \centering
  \begin{threeparttable}
 \caption{Critical amplitudes of the surface tension for the LJ fluid according to Eq.~\eqref{critamplit_st_relat}.}\label{tab_stamplit}
  \begin{tabular}{ccccccc}
  \hline
  \hline
  &\multicolumn{2}{c}{Ising}&\multicolumn{2}{c}{LJ-fluid, Eq.~\eqref{critamplit_st_relat}}&\multicolumn{2}{c}{LJ-fluid} (simulations)\\
  \hline
  &D=2&D=3&D=2&D=3&D=2&D=3\\
  &&&&&&\\
  &&&&&&2.31\tnote{c}\\
   $s^{(0)}$&4\tnote{a}&
  6.77\tnote{b}&4/3&2.82& $\approx 1.32$\tnote{g} &2.22\tnote{d}\\
  &&&&&&$2.60\pm 0.04$\tnote{e}\\
  &&&&&&$2.94\pm 0.05$\tnote{f}\\
\hline
\hline
\end{tabular}
\begin{tablenotes}\footnotesize
\item[a] Onsager's solution\hphantom{Ref.~\cite{crit_surftensionlj_jcp1995}}
\item[d] Ref.~\cite{eos_miefluid_physlett2008}\hphantom{Onsager's solu} \item[g] Ref.~\cite{liq_2dsurftens_jcp2009}\\
\item[b] Ref.~\cite{crit_surftension_fisher_physa1996},\cite{crit_surftensisingmodmc_jphysfr1993}\hphantom{Onsager's solu}
\item[e] Ref.~\cite{liq_surftenslj3d_jcp2000}\\
\item[c] Ref.~\cite{crit_surftensionlj_jcp1995}\hphantom{Onsager's solution\hspace{0.55mm}\,}
\item[f] Ref.~\cite{liq_surftens_jcp2005}\\
\end{tablenotes}
\end{threeparttable}
\end{table}

Analogous relations can be derived for the critical amplitudes of any other thermodynamic quantity according to Eq.~\eqref{projtransfr_my}. For example, for the the binodal density amplitude $B^{(0)}$, which is defined as $n_{l}-n_{g} =2\,n_c\,B^{(0)}|\tau|^{\beta}$, we obtain:
\begin{equation}\label{critamplit_dens}
 B^{(0)}_{LJ} = (1+z)^{\beta}\,B^{(0)}_{latt}
\end{equation}
Corresponding estimates for the LJ fluid amplitude $B^{(0)}$ in 2D and 3D cases are presented in Table~\ref{tab_bamplit}.
\begin{table}[hbt!]
  \centering
  \begin{threeparttable}
 \caption{Critical amplitude of the density $B^{(0)}$}\label{tab_bamplit}
  \begin{tabular}{ccccc}
  \hline
  \hline
  &\multicolumn{2}{c}{Ising}&\multicolumn{2}{c}{LJ-fluid, Eq.~\eqref{critamplit_dens}}\\
  \hline
  &D=2&D=3&D=2&D=3\\
  &&&& \\
  $B^{(0)}$&1.1\tnote{a}&
  1.69\tnote{b}&1.14&1.93\\
  \hline
\hline
\end{tabular}
\begin{tablenotes}\footnotesize
\item[a] Onsager's solution\hphantom{Ref.~\cite{crit_3disingmc_jmathphys1996}}
\item[c] Ref.~\cite{crit_surftensionlj_jcp1995}\\
\item[b] Ref.~\cite{crit_3disingmc_jmathphys1996}\hphantom{Onsager's solution}
\item[d] Ref.~\cite{eos_miefluid_physlett2008}
\end{tablenotes}
\end{threeparttable}
\end{table}

\section{Effective width of interface}\label{sec_width}
The notion of the width $\Delta$ of the interface along with the surface tension is an important integral characteristic of the phase coexistence and the corresponding inhomogeneous state. Close to the critical point it diverges as fast as the correlation length $\xi$ of the system. Yet the very definition of the interfacial width varies in different studies, since there is an arbitrariness in choosing the quantitative measure of where the surface begins \cite{liq_surfacethickness_physa1978}. In \cite{eos_zenosurftensus_jphchemc2016} it was shown how it is possible to define this quantity in an intrinsic manner directly from the Ornstein-Zernike relation, and the following expression for the surface tension of the lattice gas was suggested:
\begin{equation}\label{st_woodmodif}
  \sigma_{latt} = \sigma_{0}\,\f{t}{\xi^{1-\eta}_{eff}}\,\left(\, 1-2\,x_{gas}\,\right)\,\ln\,\f{1-x_{gas}}{x_{gas}}
\end{equation}
This expression is based on the Bragg-Williams approximation for the Woodbury's eigenvector of the bulk representation for the surface tension of the lattice gas \cite{crit_surftenslatticegas_jcp1969}. In terms of the magnetization $0\le M(t)\le 1$ of the Ising model:
\begin{equation}\label{st_woodmodif_magnet}
  \sigma_{latt} = \sigma_{0}\,\f{t}{\xi^{1-\eta}_{eff}}\,M(t)\,\ln\,\f{1+M(t)}{1-M(t)}
\end{equation}
In this paper we view the characteristic length $\xi_{eff}$ as the effective thickness of the interface in the units of the lattice spacing. Below we will show that this thickness is proportional to correlation length $\xi$, using the Trietzenberg -- Zwanzig formula for the surface tension.
Note, that introducing the length scale factor $\xi_{eff}$ in Eq.~\eqref{st_woodmodif} is important for the surface tension to have the correct critical asymptotic behavior, as the original Woodbury result fails to reproduce the correct critical behavior. The density dependent part of the expression \eqref{st_woodmodif} is due to the Bragg-Williams mean-field approximation \cite{crit_surftenslatticegas_jcp1969}. The amplitude factor $\sigma_0$ can be fixed by the the low temperature asymptote in dimensionless form:
\begin{equation}\label{st_lowtemp}
  \sigma_{latt}(t) \to 2\,,\quad \xi_{eff}\to 1\,,\quad t\to 0
\end{equation}
Here we imply that when the temperature decreases the interfacial width tends to its minimal value, which is the lattice spacing.
For the lattice model with the nearest neighbor interactions this results in $\sigma_{0} = 1/4$, since $x_{gas}\to e^{-8\,J/t}$ as $t \to 0$.

By applying the inverse transformation \eqref{projtransfrinv_my} and the relation \eqref{st_iso} to the expression \eqref{st_woodmodif} for the surface tension of the lattice gas, we obtain the following expression for the surface tension of the LJ fluid:
\begin{equation}\label{st_real}
  \sigma_{LJ} = \frac{\sigma_0}{4\,\xi_{\text{eff}}^{1-\eta}}\,\f{t_{c}}{z}\f{\,T/T_*}{\left( 1-T/T_*\right)^2}\,\f{n_{l}-n_{g}}{n_*}\,
  \ln{\f{n_{l}}{n_{g}}}
\end{equation}

\begin{figure}[hbt!]
  \centering
  \includegraphics[scale=0.65]{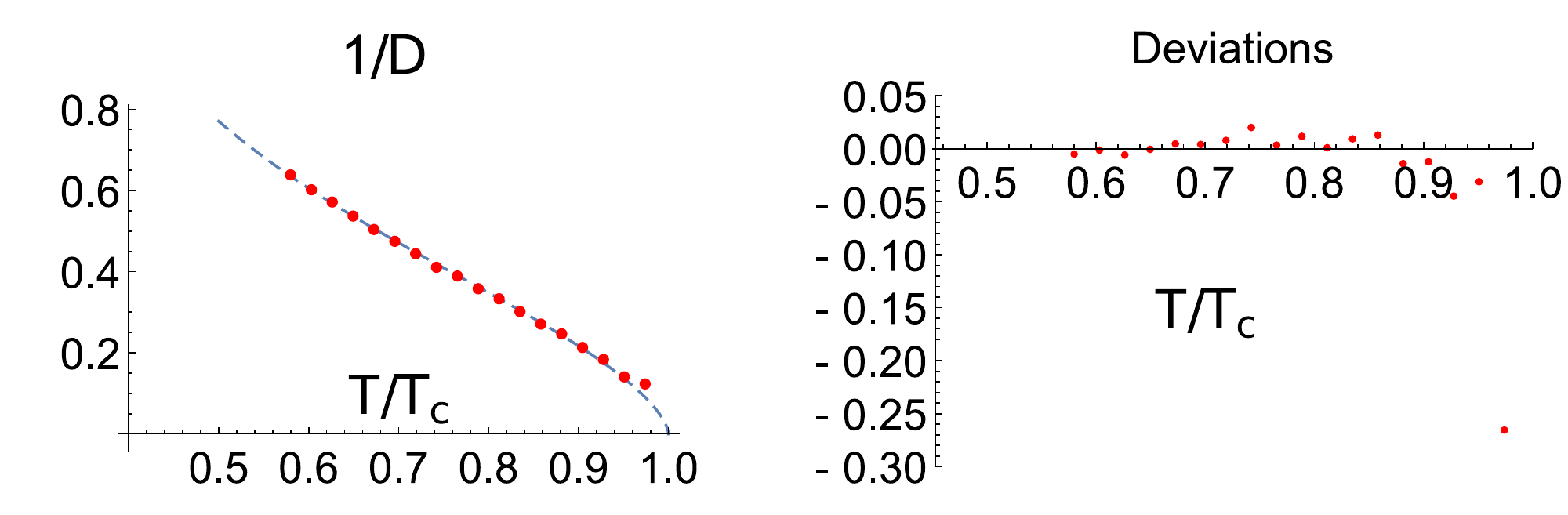}
  \caption{(left) Inverse interfacial width as a function of the temperature. The curve represents the best fit of Eq.~\eqref{xi_simplescaling} with $a\approx 0.173 $ to the literature data. The points represent the data obtained in the molecular simulations \cite{liq_surfacetension_molphys2006}. (right) Deviation of the data from the fit.}\label{fig_xiefflj3d}
\end{figure}

To demonstrate that the characteristic length $\xi_{eff}$ introduced in \eqref{st_woodmodif} is proportional to the correlation length $\xi_{latt}$ of the lattice model we use the data for the effective interfacial width $\Delta$ of LJ fluid  interface, obtained in molecular simulations \cite{liq_surfacetension_molphys2006}. We follow the authors in the assumption that the spatial density profile has the form of the hyperbolic tangent:
\begin{equation}\label{Dtanh}
  n(z) = n_{d} - \frac{n_{l}-n_{g}}{2}\,{\rm tanh}\left(\frac{2\,z}{\Delta}\right)
\end{equation}
To relate the value of the interfacial width $\Delta$ with the correlation length of the Ising model we use a simple scaling transformation\cite{eos_zenosurftensus_jphchemc2016}:
\begin{equation}\label{xi_simplescaling}
  \Delta(T) = \xi_{Is}(t(T))/a_{\Delta}
\end{equation}
and adjust the parameter $a$ to find the best fit to the simulation data \cite{liq_surfacetension_molphys2006}. We also use the relation \eqref{projtransfr_my} between the temperature parameters of the Ising model and the LJ fluid for $D=3$ with $z=1/2$.
The result is shown in Fig.~\ref{fig_xiefflj3d} with the best fit value $a_{\Delta} = 0.173$. The figure also shows the relative deviations of the data from the fit. These deviations grow when the temperature approaches the region, where fluctuations are significant and the correlation length increases essentially. In this region truncating the interaction potential leads to large errors.

In addition, to relate $\xi_{LJ}$ with $\xi_{Is}$ we can use the universality of the Fisk-Widom ratio for the critical asymptotic of the surface tension \cite{crit_surfacetension_fiskwidom_jcp1968}:
\begin{equation}\label{st_fiskwidomratio}
W_{-} = \lim\limits_{t\to t_c - 0}\,\frac{\sigma(t)}{t}\,\xi(t)^{d-1}
\end{equation}
where $\xi$ is the correlation length. Using the exact Onsager result in 2D that $W_{-} = 0.310$, together with Eq.~\eqref{st_woodmodif} and Eq.~\eqref{st_fiskwidomratio} we can derive the following estimate:
\[\xi_{eff}(t)/\xi(t)\to 1.515, \quad t\to t_c\]
Indeed, using Eq.~\eqref{critamplit_st_relat} we obtain the following relation between the correlation length critical amplitudes $\xi^{(0)} = \xi\,|\tau|^{\nu}\,,\,\,\tau\to 0$:
\begin{equation}\label{xi_fwrelat}
  \frac{\xi^{(0)}_{Is}}{\xi^{{(0)}}_{LJ}} = \left(\frac{s^{(0)}_{LJ}}{s^{(0)}_{Is}}\frac{t_c/J}{T_c/\varepsilon}\right)^\frac{1}{D-1} = \frac{D+6}{6}\,\left(\frac{t_c/J}{4\,T_c/\varepsilon}\right)^\frac{1}{D-1}
\end{equation}

In \cite{eos_zenosurftensus_jphchemc2016} it was shown that the phenomenological  expression \eqref{st_real} is in agreement with the basic Trietzenberg -- Zwanzig formula for the surface tension \cite{liq_surfacetensionzwanzig_prl1972}:
\begin{equation}\label{st_trietzwan}
\sigma = T\,\iint d\,n(z_1)\,d\,n(z_2)\,
K_2(z_1,z_2)\,.
\end{equation}
where
\begin{equation}\label{k2}
K_2(z_1,z_2) = \frac{1}{4}\,\int d^{D-1}\boldsymbol{\rho} \,\, \rho^2\,C_2\left(\, z_1,z_2;\rho\,\right)\,,
\end{equation}
while $\boldsymbol{\rho} = (x,y)$ is the vector along planar interface and $C_2$ is the direct correlation function for corresponding inhomogeneous state \cite{book_hansenmcdonald}. Using this expression it is easy to show \cite{eos_zenosurftensus_jphchemc2016} that  Eq.~\eqref{st_real} has the following asymptotic behavior
\begin{equation}\label{surftens_asympt}
  \sigma \propto \frac{(n_{l}-n_{g})^2}{\xi_{eff}^{1-\eta}} \propto   |\tau|^{2\beta + \nu(1-\eta)} =  |\tau|^{(D-1)\,\nu}\,,\quad
\end{equation}
with $\xi_{eff}\propto \xi$.

In order to extend Eq.~\eqref{st_real} to a broad temperature interval one needs some model expression for $\xi_{eff}$. In the framework of the global isomorphism the only reasonable choice is to require $\xi_{eff}$ to be proportional to the correlation length of the corresponding Ising model $\xi_{Is}$. The results for the interfacial width shown in Fig.~\ref{fig_xiefflj3d} support this choice.

\begin{figure}
  \centering
  \includegraphics[scale=0.48]{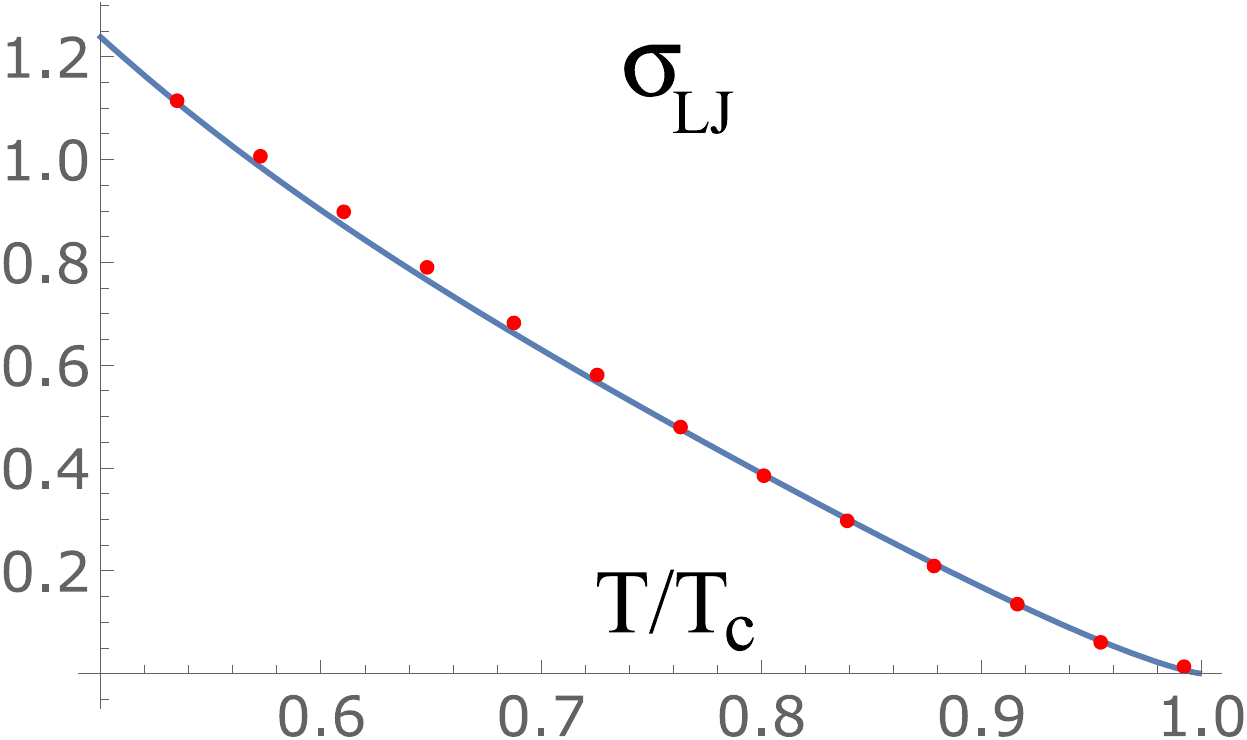}\,\,
  \includegraphics[scale=0.53]{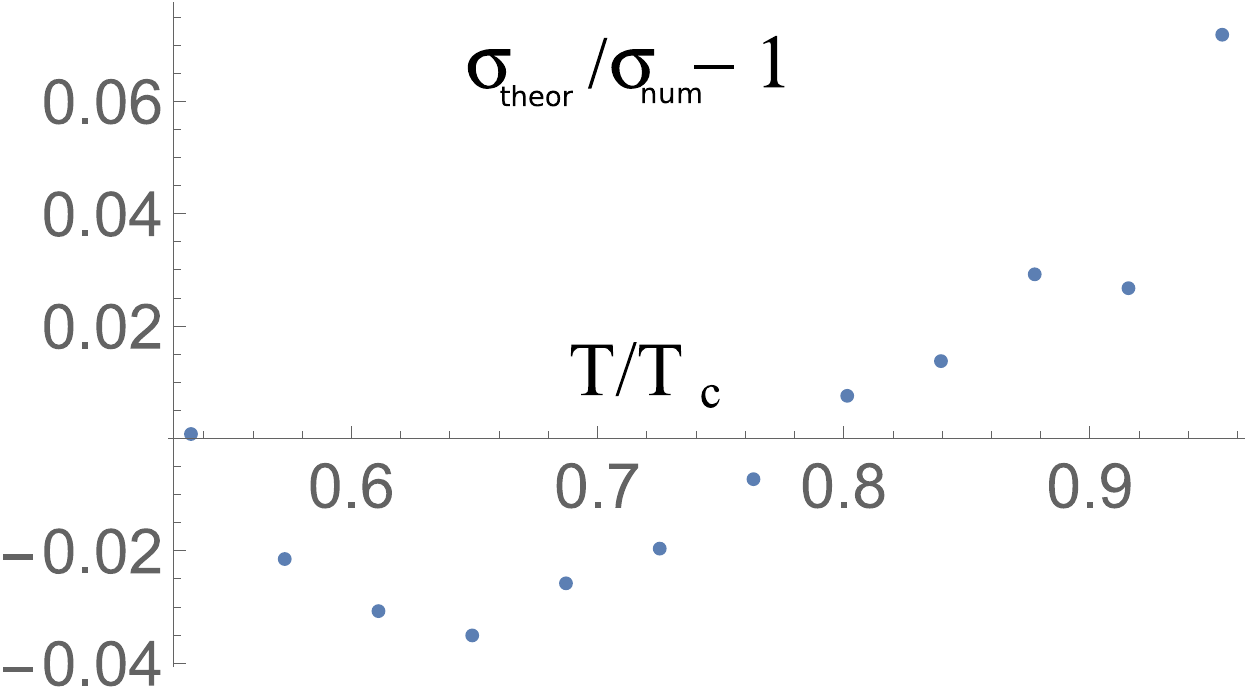}
  \caption{(left) Surface tension of the LJ fluid as a function of the temperature according to \eqref{st_real}. Points represent the data of molecular simulations \cite{eos_ljfluidinterface_jcp2009}. (right) The relative deviation of the theoretical prediction from the numerical data for various temperatures.}\label{fig_stlj3d}
\end{figure}

We next use the numerical data for the magnetization of the 3D Ising model \cite{crit_3disingliufisher_physa1989} to calculate the surface tension. We choose these data, as the system studied there has the critical temperature approximately $1.31$, which is close to the global isomorphism estimate of the critical temperature $T_{c}= 4/3$. Furthermore, the system studied in these simulations \cite{eos_ljfluidinterface_jcp2009} has the critical exponent $2\nu = 1.26$, which is close to the value $2\nu = 1.25$ of the system studied in the other simulations \cite{crit_3disingsurface_physa1993} which we used to obtain the correlation length $\xi(t)$ of 3D Ising model. The result is shown in Fig.~\ref{fig_stlj3d} along with the molecular simulation data of Galliero et al.~\cite{eos_ljfluidinterface_jcp2009}. As in Fig.~\ref{fig_xiefflj3d} the difference between the theory and molecular simulations grows as the temperature gets closer to the critical point. However, outside the fluctuation region the difference does not exceed $5\%$. Note, that the surface tension curve is fitted with the only one fitting parameter $a$, which in the present case is equal $a\approx 0.3$. No other fitting parameters were introduced to match the data of molecular simulations. In particular, the critical amplitude $s^{(0)}_{LJ}\approx 2.82$ is obtained from the Ising model. In fact this scaling factor is the only fitting parameter of the theory which has clear physical meaning and can be fixed using universal Fisk-Widom ratio. Another way of ramification of the theory is to go beyond the simplest Bragg-Williams approximation for the
Woodbury's eigenvector.

%
\section*{Conclusion}
In this paper we have applied the global isomorphism approach to relate the surface tension of the Ising model and the Lennard-Jones fluid. We used the approach proposed in \cite{eos_zenosurftensus_jphchemc2016} where it was shown that the effective interfacial width is proportional to the correlation length of the Ising model by a scaling factor $a$. In the present form of the theory its value depends on the density and is determined by the approximation used for calculation of the Woodbury's eigenvector \cite{crit_surftenslatticegas_jcp1969}. For the purposes of the paper we used this factor as the free adjusting parameter. It is the only fitting parameter of the theory which has clear physical meaning and can be fixed using universal Fisk-Widom ratio.

We have derived the relations between critical amplitudes of the surface tension of the Ising model (lattice gas) and the LJ fluid. The estimates based on this relation belong to the range calculated in the numerous computer simulations for the Lennard-Jones fluid.

Our approach is valid equally well both in 2D and 3D cases. The case of 2D geometry is especially complicated because existing theoretical approaches fail to reproduce correctly both the binodal and the surface tension data. The results, which we have obtained, clearly demonstrate that the global isomorphism is self-consistent and a useful approach. It provides the information about liquid-vapor phase coexistence of the Lennard-Jones fluid on the basis of the knowledge about lattice models. It seems natural that the fluid systems with the Mie-type as well as Yukawa and the square well potentials, where the linearities of the binodal diameter and the Zeno-line are observed, can also be mapped onto isomorphic lattice model with symmetrical binodal.
\section*{Acknowledgement}
This work is partially supported by Ministry of Education and Science of Ukraine, grant
No~0115U003214. V.K. also acknowledges Mr.~K. Yun for the support of the research. The help of Mrs. S.~Bogdan in reviewing the text of the manuscript is kindly appreciated.

\section*{References}
%

\end{document}